\documentstyle[preprint,aps,epsf]{revtex}
\newcommand{\beq}{\begin{equation}}
\newcommand{\eeq}{\end{equation}}
\def\bea{\begin{eqnarray}}  \def\eea{\end{eqnarray}}

\begin{document}

\begin{center}
\vspace*{1 truecm}
{\Large \bf 
Associated multiplicity to high $p_T$ events and percolation of color sources}\\[8mm]
{\bf  L. Cunqueiro$^1$, J. Dias de Deus$^2$, and C. Pajares$^1$}
\vskip 8 truemm

{$^1$ Instituto Galego de F\'{\i}sica de Altas Enerx\'{\i}as and Departamento de F\'{\i}sica de Part\'{\i}culas, Universidade de Santiago de Compostela, 15782 Santiago de Compostela, Spain.}
\vskip 3 truemm
{$^2$ CENTRA, Instituto Superior T\'ecnico, 1049-221 Lisboa, Portugal}

\vskip 1.0truecm
{\Large {\bf Abstract}}
\end{center}
\begin{quotation}
We show that the multiplicity distribution associated to high $p_T$ events is given in terms of the total multiplicity distribution in an universal way due to the fact that these events are self-shadowed. In particular, the mean associated multiplicity is related to the fluctuations on the multiplicity distribution.
In the framework of percolation of strings these fluctuations are related to clustering and as a result the associated multiplicity presents a peculiar dependence on centrality which can be experimentally tested.
\end{quotation}
\vskip 1.0truecm


\vspace{5cm}

\newpage

In the last few years very interesting phenomena related to high $p_T$ physics have been observed at RHIC experiments \cite{1,4}, namely a strong supression of inclusive high $p_T$ hadron production in Au-Au central collisions compared to the scaling with the number of binary nucleon-nucleon collisions. The data also show the disappearence of back to back jet-like hadron correlations in Au-Au collisions, contrary to what is observed in d-Au and p-p collisions, and a peculiar behaviour of the fluctuations in transverse momentum and multiplicity \cite{5,6,7,8} with a maximum at a certain centrality. In order to explore further the physical phenomena involved\cite{9}, different correlations related to  high $p_T$ events are being studied.
In this paper we show that the single mean associated multiplicity to high $p_T$ events carries valuable information on multiplicity correlations.
In the framework of percolation of colour sources, we predict a peculiar behaviour of the dependence of the associated multiplicity on the number of participants, with a maximum at a certain centrality related to the maximum in the number of clusters with different colour sources.

Particle production in hadron -hadron , hadron-nucleus and nucleus-nucleus is usually considered as a superposition of elementary particle emitting collisions. Let us consider events of a certain type C produced in an elementary collision. If the superposition of any number of events of type C as well as their superposition with any number of events not satisfying C also does satisfy C, these events are self-shadowed \cite{10}. Indeed, in hadron-nucleus collisions, the inelastic cross section can be written as
\beq
\sigma^{hA}(b)=\sum_{n=1}^{A}{A\choose n}(\sigma T(b))^n(1-\sigma T(b))^{A-n}\eeq
where we can write
\beq
(\sigma T(b))^n=\sum_{i=0}^n{n\choose i}(\sigma_C)^i(\sigma_{NC})^{n-i}T(b)^n\eeq
being $\sigma_C$ and $\sigma_{NC}$ the elementary nucleon-nucleon cross sections for events of type C and for the rest of events respectively. The final cross section for events of type C must count at least one elementary $\sigma_C$ in the sum, therefore
\beq
\sigma_C^{hA}(b)=\sum_{n=1}^{A}{A\choose n}\sum_{i=1}^n{n\choose i}\sigma_C^i\sigma_{NC}^{n-i}T(b)^n (1-(\sigma_C+\sigma_{NC})T(b))^{A-n}=1-(1-\sigma_CT(b))^A.\eeq
Formula (3) shows that C-events are self-shadowed. Similar considerations can be done for hadron-hadron and nucleus-nucleus collisions\cite{11}. There are many different self-shadowed events, for instance non-diffractive, annihilation, or high $p_T$ events. In all of them, depending on wether $\sigma_C$ is small or large, $\sigma_C^{hA}$ behaves like A or $A^{\frac{2}{3}}$ respectively. 

If we denote by $\alpha_C$ the probability of event C to occur in an elementary collision, one can trivially write
\beq
N(\nu)=\sum_{i=0}^{\nu}{\nu\choose i}(1-\alpha_C)^{\nu-i} \alpha_C^i N(\nu).
\eeq
and
\beq N_C(\nu)=\sum_{i=1}^{\nu}{\nu\choose i}(1-\alpha_C)^{\nu-i} \alpha_C^i N(\nu)\simeq\ \nu \alpha_C N(\nu)\eeq
\beq N_{NC}(\nu)=(1-\alpha_C)^{\nu} N(\nu) \simeq\ (1-\nu \alpha_C) N(\nu)\eeq
where $N(\nu)$, $N_C(\nu)$ and $N_{NC}(\nu)$ stand for the total number of events, the total number of events of type C and the total number of events not of type C produced in $\nu$ collisions, respectively. The last equalities of eqs.(5) and (6) hold in the limit of small $\alpha_C$.
Since
\beq
\sum_{\nu} N(\nu)=N \eeq
\beq \sum_{\nu} \nu N(\nu) = <\nu> N \eeq
\beq \sum_{\nu} N_{c}(\nu)=\sum_{\nu} \alpha_{C}\nu N(\nu) = N_C\eeq
being $N$ the total number of events and $N_C$ the total number of events of type C, we then have that the probability distribution for C events in $\nu$ collisions is 
\beq P_C(\nu)=\frac{\alpha_C \nu N(\nu)}{\sum_{\nu} N_C(\nu)}=\frac{\nu N(\nu)}{<\nu> \sum_{\nu} N(\nu)}=\frac{\nu P(\nu)}{<\nu>}.\eeq

In nucleus-nucleus collisions the dispersion, D, of the total distribution is related to the elementary nucleon-nucleon dispersion, d, and to the multiplicity, $\overline{n}$, by the equation
\beq
\frac{D^2}{<n>^2}=\frac{<\nu^2>-<\nu>^2}{<\nu>^2}+\frac{d^2}{<\nu> \overline{n}^2}.\eeq
Since $\nu$ is very high in nucleus-nucleus collisions, the second term of eq.(11) can be neglected. The same kind of approximation is valid for higher moments, what lets us to extend eq.(10) to the multiplicity distribution\cite{12,13}
\beq
P_C(n)\simeq \frac{n P(n)}{<n>} .\eeq

Notice that the right hand side of eq.(12) is independent of C. Eq.(12) has been checked in high energy pp collisions for the multiplicity associated to $W^{\stackrel{+}{-}}$ and $Z^0$ production, and also for the associated multiplicity distribution to jet production and the corresponding one to annihilation\cite{13}. In nucleus-nucleus collisions, data of ISR experiments on events with $p_T\geq 3GeV/c$ produced in $\alpha-\alpha$ collisions show that the associated multiplicity distribution satisfyes eq.(12)\cite{14}. Experimental data on the associated  distribution for high $p_T$ events from RHIC would be welcome in order to check eq.(12).

From eq.(12) we have 
\beq
\frac{<n>_C}{<n>}=1+\frac{D^2}{<n>^2} .
\eeq
 Therefore the mean associated multiplicity for high $p_T$ events can give us 
information on multiplicity fluctuations. In the framework of the percolation of strings\cite{15}, $\frac{D^2}{<n>^2}$ is given by\cite{16,17} 
\beq
\frac{D^2}{<n>^2}=\frac{<N^2>-<N>^2}{<N>^2}+\frac{1}{<n>}\equiv\frac{1}{k}+\frac{1}{<n>}\eeq
where N is the number of strings per cluster. At low density there is no overlapping of strings and every cluster contains only one string. This means that the fluctuations in the number of strings per cluster in eq.(14) vanish. As the density increases, more and more clusters with different number of strings are formed. This implies an increase in the first term on the r.h.s of eq.(14) up to a critical density. Over this density, there is no formation of new clusters and essentially only one cluster containing all the strings is formed. In this asymptotic regime, again the first term on the r.h.s of eq.(14) vanishes. ($\frac{1}{<n>}$ decreases with density very fast.)

Eqs.(10) and (12) have been obtained assuming independent superposition of elementary interactions. This is not strictely true when strings interact forming clusters, however we believe that such effects are not relevant in the present context. In fact, the multiplicity distribution P(n) can be obtained from the distribution on the number of clusters of N strings, W(N), and the distribution of the decay of each cluster P(N,n) \cite{16,17}
\beq
P(n)=\int_{0}^{\infty}dN W(N) P(N,n).\eeq
One good aproximation for the multiplicity distribution is to use the negative binomial distribution
\beq
P(n)=\frac{\gamma^k}{\Gamma(k)n!}\frac{\Gamma(n+k)}{(1+\gamma)^{n+k}}\eeq
where 
\beq
\frac{1}{k}=\frac{<N^2>-<N>^2}{<N>^2}.\eeq

The negative binomial distribution is obtained taking for P(N,n) the Poisson distribution and for W(N), the gamma distribution
\beq
W(N)=\frac{\gamma}{\Gamma} (\gamma N)^{k-1} \exp{(-\gamma N)}.\eeq
Then, instead of eq.(13) we obtain
\beq
\frac{<n>_{C}}{<n>}=\frac{1+\frac{1}{k}+\frac{1}{<n>}+\frac{k}{<n>}}{1+\frac{k}{<n>}}.\eeq
If $k<<<n>$ we are back to the independent-string regime and we can simplify expression (19) to
\beq
\frac{<n>_{C}}{<n>}=1+\frac{1}{k}\eeq
which shows that the maximum occurs for the minimum of k.

In fig. 1 we plot $\frac{<n>_C}{<n>}$ as a function of the number of participants for Au-Au collisions at $\sqrt{s}=200 GeV$. We take the values of $\frac{<N^2>-<N>^2}{<N>^2}$ from reference \cite{16}. We use eq.(20) which equals eq.(14) for number of participant nucleons greater than 12 ($Np>12$).
The fluctuations rise up to a maximum value and then decrease.

In conclusion, the dependence on centrality of the associated multiplicity in high $p_T$ events reflects the dependence on centrality of multiplicity fluctuations. 
The behaviour predicted in the framework of percolation of strings can be easily checked experimentally.

We thank N.Armesto for discussions. This work has been done under contracts FPA2002-01161 of CICYT of Spain and PGIDIT03PXIC20612 PN from Galicia.

\begin{figure}
\centering\leavevmode
\epsfxsize=6.5in\epsfysize=6.5in\epsffile{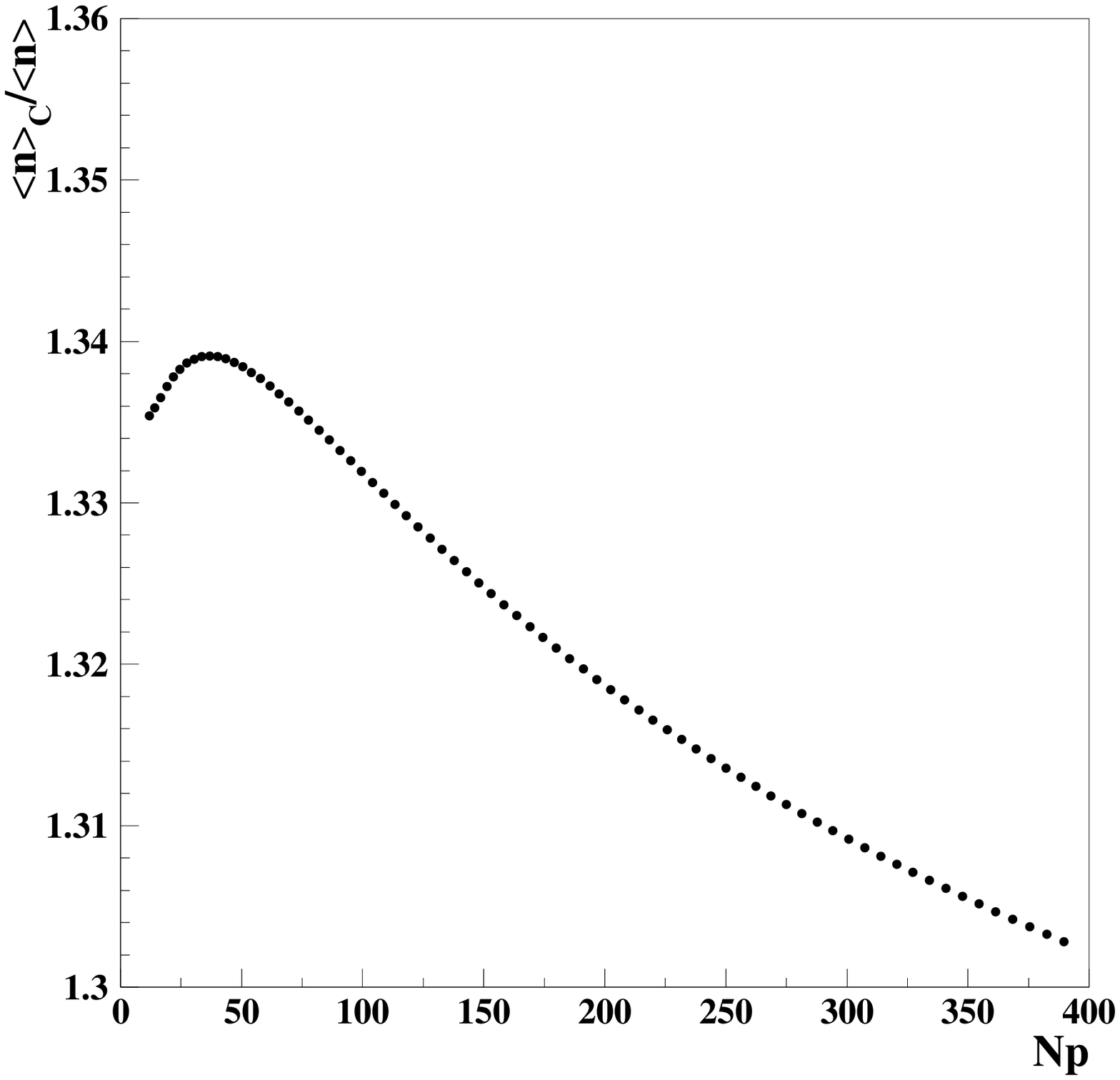}
\vskip 0.5cm
\caption{Our results for the mean multiplicity associated to rare events with respect to mean non restricted  multiplicity in Au-Au collisions at 200AGev.}

\label{fig1}
\end{figure}


\begin{thebibliography}{17}



\bibitem{1}
K.~Adcox {\it et al.}  [PHENIX Collaboration],
[arXiv:nucl-ex/0410003].

\bibitem{2}
J.~Adams {\it et al.}  [STAR Collaboration],
[arXiv:nucl-ex/0501009].


\bibitem{3}
I.~Arsene {\it et al.}  [BRAHMS Collaboration],
[arXiv:nucl-ex/0410020].

\bibitem{4}
B.B ~Back {\it et al.}  [PHOBOS Collaboration],
[arXiv:nucl-ex/0410022].


\bibitem{5}
K.~Adox {\it et al.}  [PHenix colaboration],
Phys.\ Rev.\ C {\bf 66}, 024901 (2002).
J.~Adams {\it et al.} [STAR colaboration]
Phys.\ Rev.\ C {\bf 68}, 044905 (2003).

\bibitem{6}
H.~Appelshauser {\it et al.}  [NA49 colaboration],
Phys.\ Lett.\ B {\bf 459}, 679 (1999).
H.~Appelshauser J.\ Phys.{\bf 630}, 5935 (2004).

\bibitem{7}
E.~G.~Ferreiro, F.~del Moral and C.~Pajares,
Phys.\ Rev.\ C {\bf 69}, 034901 (2004).

P.~Brogueira and J.~Dias de Deus, Acta.Phys.Pol. B {\bf 36}, 07 (2005).
 M.~A.~Braun, F.~del Moral and C.~Pajares, Eur.Phys.J C {\bf 21}, 557 (2001).
 

\bibitem{8}
L.~Cunqueiro, E.~G.~Ferreiro,  F.~del Moral  and C.~Pajares, 
[arXiv:hep-ph/0505197].



\bibitem{9}
M.~Gyulassy and L.~Maclerran, Nucl.Phys.A {\bf C750}, 30 (2005). 



\bibitem{10}
R.~Blankenbecker, A.~Capella, J.~Tran Thanh Van, C.~Pajares and A.V. Ramallo
, Phys.Lett.B {\bf 10}, 106 (1981).
 C.~Pajares and A.V. Ramallo, Phys.Lett. B {\bf107}, 373 (1981). 


\bibitem{11}
C.Pajares and A.V.Ramallo, Phys.Rev.D {\bf 31}, 2800 (1985).


\bibitem{12}
J.~Dias de Deus, C.~Pajares and C.A.Salgado, Phys.Lett. B {\bf 407}, 335 (1997),
Phys. Lett. B {\bf 409}, 474 (1997).

\bibitem{13}
J.~Dias de Deus, C.~Pajares and C.A.Salgado, Phys.Lett. B {\bf 408}, 417 (1997).

\bibitem{14}
M.~Faessler, Phys. Rep. {\bf 115}, 1 (1984). 


\bibitem{15}
N.~Armesto, M.~A.~Braun, E.~G.~Ferreiro and C.~Pajares, Phys.Rev.Lett {\bf 77},
3736 (1996). M.~Nardi, H.~Satz, Phys.Lett. B {\bf 442}, 14 (1998).


\bibitem{16}
J.~Dias de Deus, E.G.Ferreiro, C.Pajares and R.Ugoccioni, Eur.Phys.J C  {\bf 40}, 229 (2005).


\bibitem{17}
J.~Dias de Deus and R.Ugoccioni, Eur.Phys.J. C to appear. C.Pajares Eur.Phys.J C to appear, [arXiv:hep-ph/0501125].


\end{thebibliography}
\end{document}